\documentclass[final]{IEEEtran}

\usepackage{amsthm,amssymb,graphicx,multirow,amsmath,xcolor,amsfonts}
\usepackage[update,prepend]{epstopdf}
\usepackage[noadjust]{cite}
\usepackage{caption}
\usepackage[latin1]{inputenc}
\usepackage{tikz}
\usetikzlibrary{arrows,calc}		
\usepackage{bbm} 
\usepackage{textcomp}
\usepackage{pdfpages}
\usepackage{booktabs}
\usepackage{tabulary}
\usepackage{multirow}
\usepackage{comment}

\include{notation}
\allowdisplaybreaks 

\renewenvironment{IEEEbiography}[1]
  {\IEEEbiographynophoto{#1}}
  {\endIEEEbiographynophoto}

\begin{document}
\title{
Federated Learning for Wireless Communications: Motivation, Opportunities and Challenges
}
\author{
Solmaz Niknam, Harpreet S. Dhillon, and Jeffery H. Reed
\thanks{Authors are with Wireless@VT, Department of ECE, Virginia Tech, Blacksburg, VA (email: \{slmzniknam, hdhillon, reedjh\}@vt.edu). The support of the U.S. NSF (Grants CNS-1564148, CNS-1814477 and CNS-1642873) is gratefully acknowledged.} 
}

\maketitle

\begin{abstract}
There is a growing interest in the wireless communications community to complement the traditional model-driven design approaches with data-driven machine learning (ML)-based solutions. While conventional ML approaches rely on the assumption of having the data and processing heads in a central entity, this is not always feasible in wireless communications applications because of the inaccessibility of private data and large communication overhead required to transmit raw data to central ML processors. As a result, decentralized ML approaches that keep the data where it is generated are much more appealing. Owing to its privacy-preserving nature, federated learning is particularly relevant for many wireless applications, especially in the context of fifth generation (5G) networks. In this article, we provide an accessible introduction to the general idea of federated learning, discuss several possible applications in 5G networks, and describe key technical challenges and open problems for future research on federated learning in the context of wireless communications.
\end{abstract}


\section{Introduction} \label{sec:intro}
Availability of unprecedented amount of data and advancements in computing and parallel processing have led to a renewed interest in machine learning (ML) across many research fields including wireless communications.
For wireless communication, the adoption of ML for system design and analysis is particularly appealing because the traditional model-driven approaches are not rich enough to capture the growing complexity and heterogeneity of the modern wireless networks. An alternate to solely utilizing mathematical analyses, such as the ones used in model-driven communication system design, is to {\em learn} these models using massive amounts of data, which is often available to the network. This is expected to result in a complete paradigm-shift in the wireless system design.

Leveraging ML and massive amount of data has also been identified and explored as a viable solution to the pressing challenges facing the communication technology industry by leading standard development organizations in the $\text{3}^{\rm{rd}}$ generation partnership project (3GPP)~\cite{ETSI2017ENI,ATIS2018Al}. For Release 16, 3GPP has started to improve the \emph{data exposure capability} by specifying how to collect and feed the data back to the network functions for their use to support data-driven ML~\cite{TR2018Automation}. In fact, by exposing more data effectively, ML can provide better data pattern differentiation.

However, managing the large-scale data to maintain the efficiency and scalability of the ML algorithms has obviously been a challenge. {In addition, in wireless networks the data is produced by and distributed over billions of devices\footnote{As per Cisco, the number of device connections is forecasted to grow to 12.3 billion by 2020~\cite{Cisco2019VNI}.}. This necessitates the need for exploring learning solutions that can efficiently handle distributed datasets. Traditional centralized ML schemes are not quite suitable for such cases because they require the data to be transferred and processed in a central entity, which may not be possible to implement in practice due to the inaccessibility of private data. Therefore, it naturally triggers the idea of the decentralized learning solutions, in which all the private data is kept where it is generated and only locally trained models are transferred to the central entity.
Moreover, decentralized ML can significantly reduce the network bandwidth and energy consumption by sending only the features of interest rather than the stream of the raw data. Another motive behind keeping the data where it is generated and performing on-device learning is to facilitate ML to respond to real time events in latency sensitive applications. The availability of small on-device computation units, such as TrueNorth and Snapdragon neural processors, paves the way for decentralized learning solutions by providing the required hardware platform.}

\emph{Federated machine learning} is an emerging decentralized approach that is particularly cognizant of the aforementioned challenges, including privacy and resource constraints. It utilizes the on-device processing power and untapped private data by performing the model training in a decentralized manner and keeping the data where it is generated. In this article, we provide easily accessible introduction to the general concept of federated ML as an extension of the original \emph{federated approach} proposed by Google recently~\cite{bonawitz2019towards,mcmahan2016commun}. We then describe the salient features of federated ML, which differentiate it from the other decentralized learning approaches. Building on this, we discuss several key applications of the federated learning framework in fifth generation (5G) networks spanning from the content popularity prediction in edge computing architecture to the use case of federated learning in 5G core network. In order to provide a concrete example, simulations have been performed on a standard dataset to demonstrate how federated learning can be utilized to predict the content popularity in a cache-enabled network for augmented reality (AR) applications. Finally, the article concludes with an extensive discussion about challenges and future research directions. These challenges are mainly related to the security, privacy and the performance of the current federated algorithm, as well as its important considerations in wireless settings.
\begin{figure}[t]
\centerline{\includegraphics[scale=0.4]{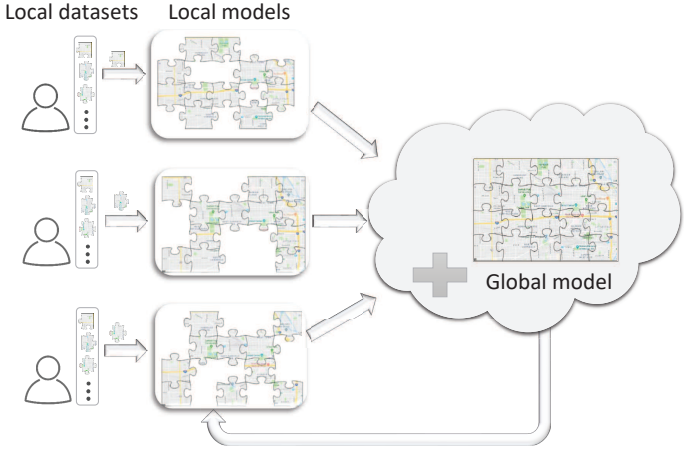}}
\caption{{An illustration of the concept of federated learning. While individually training of each local learner over its \emph{limited} dataset leads to partial models, by collaborative training, a comprehensive model can be achieved.}}
\label{fig:BigPic_FL}
\centering
\end{figure}
\section{Preliminaries and Overview}  \label{sec:Concept}
%
Recently introduced by Google, federated learning is a decentralized learning approach where training is performed over a \emph{federation} of distributed learners. It is essential to distinguish the decentralized inference approaches with centralized training from the concept of federated ML where \emph{decentralized training} is performed for decentralized inference. The objective of this approach is to keep the training dataset where it is generated and perform the model training locally at each individual learner in the federation. After training a local model, each individual learner transfers its local model parameters, instead of raw training dataset, to an aggregating unit. The aggregator utilizes the local model parameters to update\footnote{{The aggregation of the local model parameters can be accomplished either synchronously or asynchronously. Readers can refer to~\cite{bonawitz2019towards} for more details.}} a global model which is eventually fed back to the individual local learners for their use. As a result, each local learner benefits from the datasets of the other learners only through the global model, shared by the aggregator, without explicitly accessing their privacy-sensitive data. While this scheme is inherently more privacy-preserving than sharing raw data, some models may still reveal information about the underlying data because of which local learners add an additional layer of protection by transferring encrypted versions of their models to the aggregator. A secure aggregation algorithm as a class of secure multi-party computation is used to aggregate the encrypted local models without the need for decrypting the models~\cite{bonawitz2016practical}. An illustration of the federated learning concept is provided in Fig.~\ref{fig:BigPic_FL}.

Several key aspects of federated learning differentiate it from the existing distributed learning schemes. One of the common assumptions of such learning schemes is that the data samples of learners are realizations of independent and identically distributed (iid) random variables. However, in the federated ML setting, different learners may be observing separate parts of the process (with possible overlaps between them), thus generating datasets that may not be representative of the distribution of the entire data. Therefore, federated learning deals with {\em non-iid} datasets of the locally-trained learners.
As an example, one can consider the task of building a high definition (HD) map for autonomous driving, where the autonomous vehicles only collect the location and sensing information related to the routes they traverse; or in the task of hand-written digits recognition where local learners have samples of different digits. Second, the datasets are \emph{unbalanced} in size. For instance, in the HD map example, the dataset collected at different autonomous vehicles may vary in size due to different environment they pass through. Last, the datasets are \emph{massively distributed} among the local learners, where the number of data samples per local learner is smaller than the total number of learners participating in the training.
These salient features of the dataset, i.e. non-iid, distributed and unbalanced training data, differentiates the federated ML framework from the other related approaches, which are discussed below.
\begin{itemize}
\item \emph{Distributed learning} schemes are the ones in which the aggregator organizes the locally collected data (usually in the form of locally trained models due to the stringent communication limitations) to provide a holistic and more accurate estimation of the parameters under study. In this form of learning, the local learners act solely as local data collectors and do not require the global model through any feedback from the aggregator. Distributed learning in wireless sensor network (WSN) for monitoring belongs to this category of learning. For instance, in temperature monitoring WSN, each sensor in the network communicates the local model trained by its dataset to the fusion center. The fusion center aggregates the local information to construct a global estimate of the temperature of the field.

\item \emph{Parallel learning\footnote{{In the ML community, it is often called distributed ML. However, we decided to use the term \emph{parallel learning}, owing to its objective which is \emph{parallelizing} the gradient computation and aggregation across multiple worker nodes, to distinguish this type of learning from the distributed learning that we previously discussed in the context of WSN networks.}}} refers to the learning schemes whose main objective is to scale up the algorithm or accelerate the learning process or both. In this type of learning, the available training set at a central parameter server is divided into subsets of data and assigned to a group of worker machines. Therefore, the datasets assigned to each worker machine have the same underlying distribution. Subsequently, the training process is performed in parallel and the parameters are fed back to the parameter server. In this setting, model parallelism\footnote{In model parallelism, the entire dataset is assigned to all worker machines. However, each machine is responsible for estimating certain model parameters.} is another way of distributing the workload compared to the data parallelism. This type of learning is performed in datacenters where the worker machines obtain data from a shared storage and hence, unlike federated learning, they will end up having samples from the same distribution. In addition, the average number of data samples per worker is way larger than the number of worker machines participating in the training process which is different from the federated setting where the data is massively distributed.

\item \emph{{Distributed ensemble learning}}, also known as committee-based learning, is a learning approach in which multiple learners (such as classifiers and regressors) are combined to improve the overall performance. In this scheme, portions of the dataset are assigned to train different models. These models are then aggregated to reduce the likelihood of choosing an insufficient one. In general, the goal of such learning methods is to learn from a mixture of experts (models) rather than improving a global model using a naturally distributed dataset through a federation of local learners with communication constraints.

\end{itemize}
\begin{table*}[]
\centering
\captionsetup{labelfont=sc,labelsep=newline,justification=centering}
\caption{\sc{{Features, design goals and applications of Federated ML and other distributed approaches.}}}
\begin{tabular*}{0.92\textwidth}{p{2.5cm}p{8cm}p{4.6cm}}
\toprule
\textbf{Scheme}      & \hspace{0.35cm}\textbf{Salient features and design goals}                                                & \hspace{0.25cm} \textbf{Example} \\ \midrule
  \textbf{Distributed learning}
     &\begin{itemize}
     \vspace{-0.25cm}
  \item The goal is to provide a holistic estimation of the parameters under study
  \item The global model is not fed back to the local learners
  \vspace{-0.25cm}
  \end{itemize}
  &\begin{itemize}
  \item Distributed learning in WSN
  \vspace{-0.25cm}
  \end{itemize} \\\midrule
\textbf{Parallel learning}    & \begin{itemize}\vspace{-0.25cm}
  \item The goal is to accelerate the learning process and scale up the algorithm
  \item Data is distributed in a iid fashion
  \item Data is not massively distributed among learners
  \item There is no communication constraint consideration
  \vspace{-0.25cm}\end{itemize}   &   \begin{itemize}
  \item Distributed learning in datacenters environment
  \vspace{-0.25cm}\end{itemize}                                                                                                                           \\ \midrule
\textbf{Ensemble learning }   &      \begin{itemize}\vspace{-0.25cm}
  \item The goal is to produce an optimal model by learning from a mixture of several types of the models
  \item Data is distributed in a iid fashion
  \item There is no communication constraint consideration
  \vspace{-0.25cm}\end{itemize}& \begin{itemize}\vspace{-0.25cm}
  \item Bagging, boosting and stacking algorithms that can be used in remote sensing, face recognition and so on.
    \vspace{-0.25cm}\end{itemize} \\ \midrule
\textbf{Federated learning}   & \begin{itemize}\vspace{-0.25cm}
  \item The goal is to perform the model training using the naturally distributed datasets over several learners
  \item The global model is fed back to the local learners for their use
  \item Data is distributed in  non-iid fashion
  \item Data is massively distributed over local learners
  \item There are communication constraints such as privacy, security, power and bandwidth limitations in accessing the data
  \vspace{-0.25cm}\end{itemize} &\begin{itemize} \vspace{-0.25cm}
  \item Edge computing and caching
  \item Autonomous driving
  \item Federated ML for spectrum management
  \item Coexistence\,\,of\,\,heterogeneous\,\,systems (For example, DSRC and c-V2X)
  \item Federated ML in 5G core network
    \vspace{-0.25cm}\end{itemize}  \\ \bottomrule
\end{tabular*}
\end{table*}

\section{Applications of federated learning for Wireless Communications} \label{sec:Application}
After introducing federated learning and describing some of its salient features, we will now elaborate on a few of its use cases in the area of wireless communications. These applications are primarily inspired from the expected applications of 5G networks.
\subsection{Edge Computing and Caching}
\begin{figure}[t]
\centerline{\includegraphics[scale=0.5]{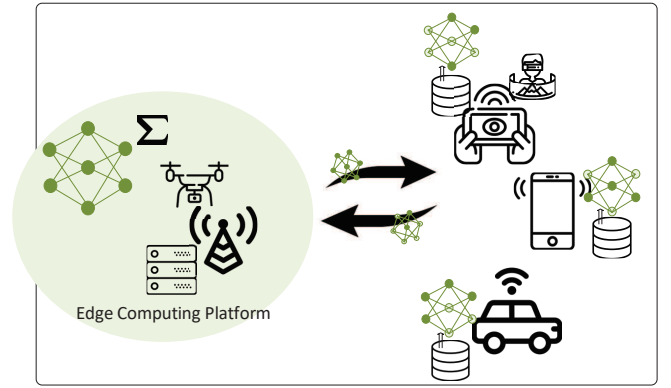}}
\caption{{ Illustration of the application of federated learning for edge computing and caching. Here, the local learners can be edge users (self-driving cars in an autonomous vehicle network or {users\textquotesingle} augmented/virtual reality headsets) and the aggregator can be an edge computing platform (such as radio base stations or unmanned aerial vehicles) in the edge network.}}
\label{fig:Edge_Comp}
\end{figure}

Content caching and data computing at the edge of the wireless network is a promising approach to reduce backhaul traffic load.
The general idea is to bring the popular content closer to the edge terminals, namely small base stations (SBSs) and access points (infrastructure caching) or even user devices (infrastructure-less caching), such that it can be conveniently accessed locally. Such a paradigm has the potential of enabling applications with stringent delay and bandwidth requirements. The success of this architecture relies on precisely determining which contents should be placed in each cache, which is an active area of research.
The approach that is usually taken in the literature is to utilize static or dynamic statistical models for content popularity identification. Unlike static models that do not capture the time varying nature of the real-time content popularity, dynamic models reflect the instantaneous popularity by considering the statistical properties of the content. {However, model-driven content popularity identification is not capable of considering multitude of factors that influence content popularity. Moreover, directly accessing the privacy-sensitive user data for content differentiation may not be possible in practice. Federated learning with the premise of utilizing the locally trained models rather than directly accessing the user data seems to be a match made in heaven for content popularity prediction in proactive caching in wireless networks (see Fig.~\ref{fig:Edge_Comp}). For instance, in AR, federated learning can be used to learn certain popular elements of the augmentations from the other users without obtaining their privacy-sensitive data directly. This {\em popular} information is then pre-fetched and stored locally to reduce the latency.} In addition, in self-driving cars, information related to traffic can be learned through other vehicles using federated learning and pre-cached in road side units.

\begin{figure}[t]
\centerline{\includegraphics[scale=0.3]{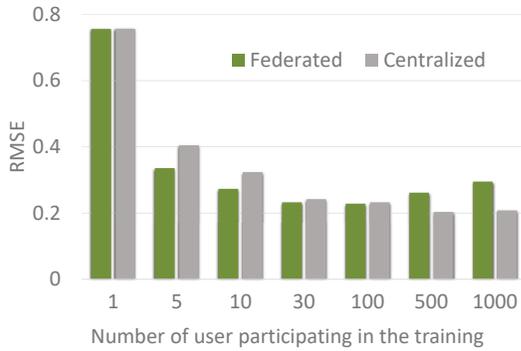}}
\caption{Comparison of the error performance of the federated learning and the baseline centralized schemes.}
\label{fig:Chart}
\end{figure}

In order to concretely demonstrate the applicability of federated learning, we have carried out simulations to predict content popularity in a cache-enabled network for AR applications. We consider a scenario where AR-enabled users hold up their device camera on a target place (such as museum, restaurant, amusement park and so on) to get more information about it.
{To reduce the latency in the AR-based demonstration and improve users experience, the popular content related to a specific place is predicted and cached proactively. However, the selection of popular content is based on search history of the users and their interaction with the content. Unfortunately, such information is private in nature and cannot be shared with the network most of the time, even though it could have significantly improved the content popularity prediction. In order to preserve the user privacy and improve the service quality at the same time, we invoke federated learning to predict content popularity based on the user-content interaction.}
We utilize AutoEncoders (AE) to predict the top contents (or rating/interaction score for the contents) that would be more appealing to the user, using the publicly available dataset MovieLens 1M. The parameters of an AE with 1 hidden layer of 128 neurons are learned by each user/learner to minimize the reconstruction error (in terms of root mean square error, or RMSE) in the federated setting. Fig.~\ref{fig:Chart} demonstrates the RMSE versus different number of users that participate in the training during each round. In addition, we considered a baseline scenario (centralized) where an AE is trained on the raw training samples obtained directly from the users rather than aggregating the individually trained models. Although implementing this in practice may not be possible because of the privacy concerns of sharing user-content interaction with the network, we consider this as a baseline case for the sake of comparison as well as demonstrating the effectiveness of federated learning. From the figure, we clearly observe that federated learning performs almost as well as the centralized scheme.
Therefore, in this case, transmitting locally trained models to the aggregator is almost as efficient as transmitting the raw data.

\subsection{Spectrum Management} \label{subsec:Handover}
The physics of propagation at millimeter wave (mm-wave) frequencies provides an opportunity to rethink the rules of spectrum access. In future 5G networks, a hybrid spectrum landscape of low and high frequencies (i.e. microwave and mm-wave bands) with different types of licensing is necessary to enable key 5G verticals. The hybrid spectrum access needs collaborative and more autonomous spectrum sharing strategies that are adapted to the environment and applications in 5G networks. However, accessing the spectrum dynamically and in a distributed manner is complicated. {The high-resolution spectrum utilization data of all radios may be required, which may not be easy to share because of privacy concerns. In fact, all radios need to share their sensory data such as spectrum occupancy data, device non-linearity information and detection of abnormal signals, such as interference. However, these data are privacy sensitive, and radios may not be willing to send out information related to their frequencies of operation.
In addition, centralized strategies, where spectrum usage information is gathered in a spectrum access database, may not always be appropriate. Not to mention that making inference on such huge amounts of data requires enormous processing power and large scale optimization that would be computationally prohibitive.} Therefore, the future of spectrum autonomy likely depends on crowd-sourced and decentralized intelligent radio networks where spectrum sharing is performed collaboratively. Federated ML, where each radio transfers its local spectrum utilization model, can be leveraged to address these issues. The aggregator utilizes the local spectrum utilization model parameters to update a global model which is eventually fed back to the individual radios for spectrum access decision. It is worth noting that the same strategy can also be used to facilitate coexistence of two wireless systems. A specific setting of current interest that can benefit from such a solution is the coexistence of dedicated short-range communication (DSRC) and cellular-connected vehicle-to-everything (c-V2X) in the same intelligent transport systems (ITS) band.

\subsection{5G Core Network} \label{subsec:Core}
\begin{figure}[t]
\centerline{\includegraphics[scale=0.45]{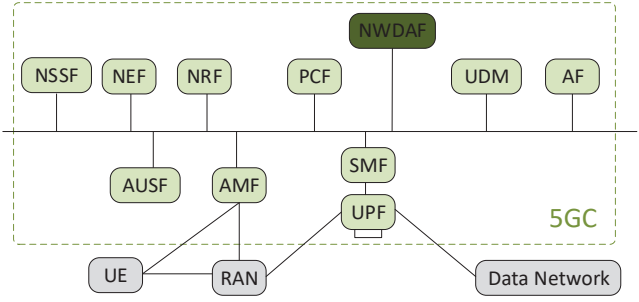}}
\caption{3GPP 5G system architecture~\cite{TR2018Automation}.}
\label{fig:5G_sys_arch}
\end{figure}

Network data analytic function (NWDAF) is a new network function defined by 3GPP to provide more data exposure capability for ML-enabled functionalities even in the core network. It provides the ability to make use of intelligent techniques in the network management system. This enables the operators to automate the network management and configuration tasks which in turn lowers the operational expenditure by reducing the human-machine interaction. In general, NWDAF is capable of connecting to any network function (NF) and utilizing any data in the core network (see Fig.~\ref{fig:5G_sys_arch}). In addition, any NF can request network analytic information.

{Now back to the federated setting, we have thus far considered that while the datasets may be based on observing different parts of the process (different sample spaces), they all contain the same feature space.
\emph{Horizontal fragmentation} is the technical term for this type of data distribution. However, there could be situations in which distributed datasets may share the same sample space but differ in feature space, namely \emph{vertically fragmented}. Inspired by the notion of data fragmentation,~\cite{Yang2019FML} has introduced vertical federated ML for vertically fragmented datasets over the federation of local learners.
It is worth mentioning that in vertical federated ML, features such as having non-iid, unbalance and massively distributed datasets are considered over feature space.
In order to understand the idea of vertically fragmented datasets completely, lets consider two datasets that cover all the subscribers of the network (and hence have the same {\em sample space}). However, they could easily differ in terms of the features. For instance, the first dataset could contain the registration and authentication information while the second could contain information related to the network slice selection for each user.}
Given such description, vertical federated learning best fits the core network structure, where each entity handles certain features of dataset related to the overall users in the network. For instance, access mobility management function\footnote{In 5G core architecture, entities are now referred to as functions to emphasize on them being virtual rather than physical entities.} (AMF) and session management function (SMF) manage mobility and session establishment (IP address allocation, traffic routing and so on), respectively. For more details on the rest of the functions, interested readers are advised to refer to~\cite{TR2018SysArch}.
Here, NWDAF can act as the global node that handles the aggregation of the user data. The datasets of the users are vertically fragmented over different entities in the core network, where each entity keeps record of a specific data feature related to all the users. Using vertical federated learning, each entity in the core network transfers its local encrypted model trained by locally collected data features rather than sending the raw data to the NWDAF entity. This can significantly alleviate the massive cybersecurity vulnerability within the network topology introduced by network function virtualization (NFV).
\section{Challenges and Future Direction} \label{sec:challenge}
Research on federated learning is still in its early stages. Despite the apparent opportunities it offers from the edge to the core networks, there exist several critical challenges in applying federated learning to wireless networks. Some of the challenges and future research directions are discussed next.

\subsection{Security and Privacy Challenges and Considerations}
Protecting privacy of the local datasets is the fundamental premise of the federated ML.
{To prevent models from revealing their data, a \emph{secure aggregation} algorithm has been proposed to aggregate the encrypted local models without the need for decrypting them in the aggregator~\cite{bonawitz2016practical}. However, the participation of a specific local learner can still be disclosed through analyzing the global model~\cite{wang2019beyond}.}
Differentially private federated algorithms have been proposed~\cite{geyer2017differentially} to provide privacy at local learner-level rather than protecting a single data sample.
However, these algorithms sacrifice the model performance or require extra computation and specific number of local learners to participate in the model training. Therefore, efficient federated algorithms that deliver models with high performance as well as privacy protection without adding computational burden are highly desirable.
{In addition, some neural network models might unintentionally memorize unique aspects of the training data~\cite{carlini2019secret}. This is in fact an important issue in case of federated learning where models are trained over sensitive user data. Given the fact that the premise of federated learning is to utilize the user data without revealing the private information, data memorization should be efficiently handled to reduce the likelihood of data disclosure in case of an attack.}

Similar to the other ML approaches, in federated learning, local models are often re-trained by the newly collected data to reflect the changes on the trained model. Therefore, an adversary can surreptitiously influence the local training datasets to manipulate the result of the model by embedding carefully designed sample to \emph{data-poison} the federated learning process. It can even threaten the model by sending gradient updates to perform \emph{model-poisoning attack}. Federated learning has been analyzed through an adversarial lens to examine the vulnerability of the learning process to the model-poisoning adversaries~\cite{bhagoji2018lens}. Poisoning resilience defense mechanisms are urgently required, as federated learning in its primary form is susceptible to such adversarial attacks.
{In addition, a curious aggregator or even a local learner can perform membership inference attacks against other local learners. In an inference attack, the attacker\textquotesingle s objective is to infer if a particular data point belongs to the training dataset~\cite{melis2019exploiting}. The repeated update of the model parameters is a key factor in boosting the accuracy of membership attacks. There are various types of inference attacks such as, parameter inference, input inference and attribute inference attacks, that can jeopardize the privacy of the local learners. Therefore, vulnerability of federated learning to these attacks and the corresponding defense mechanisms should be investigated, as well.}

\subsection{Challenges and Considerations Related to the Algorithm}
{Similar to almost every decentralized algorithm, one of the essential considerations of federated learning is the convergence of the algorithm under limited communication and computation resources. Theoretical analysis on the convergence bounds of the gradient descent based federated learning for convex loss functions has been carried out in~\cite{wang2018adaptive}. Analytical evaluations on the circumstances under which the algorithm converges for non-convex loss functions are beneficial as well, as in some models including deep neural networks, the natural objective of the model is to learn a non-convex function.}

Furthermore, considerations such as optimum number of local learners to participate in the global update, grouping of the local learners, and frequency of local updates and global aggregation, that induce trade-off between model performance and resource preservation, are application-dependent and worth investigation.
In addition, for some models such as federated deep neural networks, even the updates might still be large in size for low-powered devices such as IoT nodes. Therefore, approaches that sparsify and compress the model parameters are computationally efficient and reduce the resource consumption.

\subsection{Challenges and Considerations in Wireless Settings}
{Owing to the limited capacity of wireless channels, information needs to be quantized before it is sent over the channel. Since local learners and the aggregator need to exchange model parameters over the wireless channel, this would give rise to the paradigm of federated learning with parameter quantization. One important consideration in such a paradigm would be the robustness of models in the presence of quantization error. Besides communication bandwidth, noise and interference are other factors that exacerbate the channel bottleneck. Robustness to these channel effects should be considered, as well.}

{Another important consideration is the convergence time. The convergence time federated learning includes not only the computation time on the local learners and the aggregator but also the communication time between them which depends on the wireless channel quality. Therefore, wireless channel quality should be considered when optimizing the frequency of the local updates and global aggregation. }

{Moreover, when learning a deep model, there are model compression techniques and sparse training approaches to reduce the complexity of the model and scale down the model parameters. These approaches are useful for devices with limited processing power to learn deep models. Therefore, there is a tradeoff between reducing the complexity and maintaining the accuracy of the model. In federated deep learning for wireless applications, communication cost and quality of the wireless channel should also be considered in the model optimization. In addition, given the time varying nature of the wireless channel, model compression can be done in an adaptive manner depending on the quality of the wireless channel.}

{Besides devices availability and their willingness to participate in the learning process, quality of the wireless channel between the global aggregator and a specific local learner will also impact its selection for training and should be considered jointly with other factors. There might be cases where a particular device is willing to contribute but its corresponding wireless channel is not strong enough to transfer the model parameters with predetermined quality, which may degrade the accuracy of the global model.}
\section{Concluding Remarks} \label{sec:Conclusion}
This article discussed the role of federated ML in addressing some of the challenges in wireless communications mainly related to the 5G paradigm. Federated ML is an emerging decentralized learning solution that tries to address the energy, bandwidth, delay and data privacy concerns in wireless communications by performing decentralized model training.
We started by providing an accessible introduction to the concept of federated learning and its salient features. We then introduced several use cases of federated learning in 5G networks, spanning from edge to the core network. Simulations have been carried out to demonstrate the applicability of federated learning to content popularity prediction in a cache-enabled network for AR applications. Our results indicate that federated learning could approach the performance of the centralized scheme in which the training is performed centrally by transferring all the data from the users to the central node (which is often not possible in practice due to privacy concerns). Numerous issues and open challenges are also discussed that require further research effort in this direction.

\bibliographystyle{IEEEtran}
\bibliography{IEEEabrv,GBbibfile}
\vskip 0pt plus -1fil
\begin{IEEEbiography}{Solmaz Niknam}
received her Ph.D. degree from Kansas State University, KS, USA in 2018. During her Ph.D., she was a recipient of the Kansas Ph.D. students Fellowship. She is currently a postdoctoral associate at Virginia Tech. Her research interests include wireless communication with emphasis on 5G mm-wave networks and ML/AI in communication.
\end{IEEEbiography}
\vskip 0pt plus -1fil
\begin{IEEEbiography}{Harpreet S. Dhillon} (S'11-M'13-SM'19) is an Associate Professor of Electrical and Computer Engineering and the Elizabeth and James E. Turner Jr. '56 Faculty Fellow at Virginia Tech. He received his B.Tech. degree from IIT Guwahati in 2008, his M.S. degree from Virginia Tech in 2010, and his Ph.D. degree from the University of Texas at Austin in 2013, all in Electrical Engineering. His research interests include communication theory, wireless networks, stochastic geometry, and machine learning. He is a Clarivate Analytics Highly Cited Researcher and a recipient of five best paper awards. He serves as an Editor for three IEEE journals.
\end{IEEEbiography}
\vskip 0pt plus -1fil
\begin{IEEEbiography}{Jeffrey H. Reed} (F'05) is currently the Willis G. Worcester Professor with the Bradley Department of Electrical and Computer Engineering, Virginia Tech. He received his B.S., M.S., and Ph.D. degrees from the University of California, Davis, CA, USA, in 1979, 1980 and 1987, respectively. He is the Founding Faculty Member of the Ted and Karyn Hume Center for National Security and Technology. In 2012, he served on the President's Council for the Advisors of Science and Technology Working Group and is currently interim director of the Commonwealth Cyber Initiative. He is the author of three books and over 200 journal and conference papers.
\end{IEEEbiography}

\end{document}